\documentclass[twocolumn,showpacs,preprintnumbers,amsmath,amssymb]{revtex4}
\usepackage{graphicx}% Include figure files
\usepackage{dcolumn}% Align table columns on decimal point
\usepackage{bm}% bold math

\begin{document}

\preprint{}

\title{What do the phase-sensitive experiments tell us?}

\author{Yunping Wang}
\email{ypwang@aphy.iphy.ac.cn}
\author{Li Lu}
\author{Dianlin Zhang}
\affiliation{Key Laboratory of Extreme Conditions Physics, Institute of  Physics \& Center for Condensed Matter Physics, Chinese Academy of Sciences, P. O. Box 603, Beijing 100080, P. R. China \\}

\date{\today}

\begin{abstract}

The phase-sensitive experiments on cuprate superconductors have told us about the symmetry of the condensate wavefunction. However, they can not determine the pairing symmetry of Cooper pairs. To describe a superconducting state, two wavefunctions are needed, condensate wavefunction and pairing wavefunction. The former describes the entirety movement of the pairs and the latter describes the relative movement of the two electrons within a pair. The $\pi$-phase shift observed in the phase sensitive Josephson measurements can not prove that the pairing state is d-wave. We present here a new explanation and predict some new observable phenomena.

\end{abstract}


\maketitle

The pairing symmetry in cuprate superconductors have been receiving great attention \cite{Levi, Anderson, Chakravarty, Cox,  Clery, Kirtley, TsueiRV, VanHarlingenRV, SigristRV, ScalapinoRV}. More and more phase-sensitive experiments on Josephson junctions have demonstrated that the superconducting condensate has a $\pi$-phase shift between (100) and (010) surfaces \cite{TsueiRV, VanHarlingenRV}. This $\pi$-phase shift has been regarded as a direct evidence of d-wave pairing in cuprate superconductors ever since \cite{Levi, Clery, Kirtley, TsueiRV, VanHarlingenRV}. However, we will make it clear that the superconducting condensate wavefunction describes the behavior of bosons in condensate while the superconducting pairing wavefunction describes the pairing state in forming a Cooper pair, and therefore these two wavefunctions should be distinguished and considered separately. The phase-sensitive experiments only tell us about the symmetry of the superconducting condensate wavefunction, they are not necessary related to the pairing state of a Cooper pair.

First let us pay attention to some inevitable paradoxes when we describe the Josephson tunneling in current theoretical frames \cite{TsueiRV, Geshkenbein, Rice}.

The $\pi$-phase shift was theoretically predicted by Geshkenbein {\em et al.} for heavy-fermion superconductors \cite{Geshkenbein}. In 1992, Sigrist and Rice proposed that the $\pi$-phase shift could also be observed in cuprate superconductors \cite{Rice}. In their picture, the basic notion for the $\pi$-phase shift was that the superconducting wavefunction  $\Psi({\bf k})$ depends on the direction of the momentum wavevector ${\bf k}$ ($|{\bf k}|\sim k_F$) of the paired electrons. The wavefunction should obey $\Psi(-{\bf k})=-\Psi({\bf k})$ for p-wave pairing superconductors, and $C_4 \Psi({\bf k})=-\Psi({\bf k})$ ($C_4$ is 90$^{\circ}$ rotation operator) for d-wave pairing superconductors. 

\begin{figure}
\includegraphics{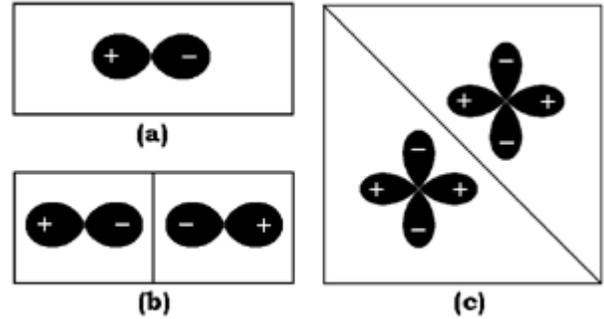}
\caption{\label{fig:paradox} paradoxes in $\pi$-phase shift for p-wave and d-wave superconductors. (a)p-wave superconductor; (b) p-wave superconductors with a barrier; (c) d-wave superconductors with a  barrier along $45^{\circ}$ direction.}
\end{figure}

We may perform a {\it gedanken} experiment for a p-wave superconductor. Imagine that a barrier is inserted into the sample (Fig.~\ref{fig:paradox}a) and makes it two p-wave superconductors with a Josephson junction in-between. The Josephson coupling requires that the superconducting wavefunction has the same sign \cite{Geshkenbein} at the two sides of the junction as shown in Fig.~\ref{fig:paradox}b. With thinning the barrier, these signs should not change. When the thickness of the barrier shrinks to zero, the two parts will return to its original single superconductor (Fig.~\ref{fig:paradox}a) but with no sign difference in wavefunction between its two ends. This is a paradox.

A paradox also occurs for d-wave superconductors. Consider a d-wave superconductor with strict tetragonal symmetry (Fig.~\ref{fig:paradox}c). Again, imagine a barrier along $45^{\circ}$ direction which divides the superconductor into two triangular superconductors. Since $\Psi({\bf k})=0$ at $45^{\circ}$ direction, there is no Josephson coupling for an ideal junction. In other words, the two triangular superconductors on the two sides are not coherent with each other. There are infinite number of $45^{\circ}$ planes in the superconductor, and every plane divides the superconductor into two parts without phase correlation. Then, how could the superconductor be a superconductor? 

\begin{table*}
\caption{\label{tab:table1} Quantum interferences of three different kinds of particle pairs}
\begin{ruledtabular}
\begin{tabular}{l|c|c|c}
					&Josephson			&Two-photon			&Interference		\\
					&Interference&Interference \cite{two-photon}&of Na atoms \cite{NaBEC}     \\ \hline
\multicolumn{4}{l} {\sl special properties}\\ \hline
~basic units					&Cooper pairs		&two-photon pairs		&atoms			\\ \hline
~constituent particles			&electrons       		&photons   			&electrons and ions	\\ \hline
~statistic property of constituent particles	&Fermi  		&Bose 			&Fermi 			\\ \hline
~exchange symmetry	of internal wavefunction&antisymmetric		&symmetric			&no 		\\ \hline
~interaction between two constituent particles	&pairing potential	& no interaction		&Coulomb potential\\ \hline
~internal states	&spairing states		&entanglement state \cite{EPR} 	&atomic states		\\ \hline
~internal relative variable	& ${\bf k}-(-{\bf k})=2{\bf k}$	&$\omega_1-\omega_2$	&electron relative position 	\\ 
\hline
\multicolumn{4}{l} {\sl common properties}\\ \hline
 \multicolumn{4}{c}{The value of internal variable is undefinable} 							\\ \hline
\multicolumn{4}{c}{Interference pattern depends on entire wavefunction of pairs}	\\
\end{tabular}
\end{ruledtabular}
\end{table*}

The key to avoid the above paradoxes is that one should not mix up the properties of the superconducting condensate with the pairing state of a Cooper pair.

A Cooper pair consists of two particles, it has double degrees of freedom compared to a single particle. We need a condensate wavefunction $\Psi({\bf R})$ to describe the boson system with the Cooper pairs as the basic units. We also need a pairing wavefunction $\psi({\bf r})$ to describe the relative movement between the two particles inside a Cooper pair.  Needless to say,  the phase-sensitive experiments measure the interference effect of the Cooper pairs.  Therefore, these experiments can only give  imformation about  the condensate wavefunction, $\Psi({\bf R})$, not the relative movement characterized by the pairing wavefunction $\psi({\bf r})$.

Interference effect is not necessarily related to the internal structure, no matter the basic units are single particles or of complex structure. In Table~\ref{tab:table1},   three different kinds of quantum interferences of particle pairs are compared. In  Na atom, what we know about the internal structure is that the outer shell electron may occupy  3s, 3p, or 4p states, etc, but we do not know the definite position of the electron. For a two-photon Einstein-Podolsky-Rosen (EPR) pair \cite{EPR}, we can never tell the difference between the frequencies of the two paired photons. Similarly, for a Cooper pair, we can never figure out the definite relative momentum $2{\bf k}$ between the two paired electrons. We may only say  that the pair is in s-wave, p-wave, or d-wave pairing state. 

Furthermore,  the internal variable of a Cooper pair is not related to the $\pi$-phase shift in Josephson tunneling experiments. If the Cooper pair composed of two electrons with ${\bf k}$ and $-{\bf k}$ along x-axis have larger tunneling probability in the x-direction than that with ${\bf k}$ and $-{\bf k}$ along y-axis  as supposed in ref.~\cite{Rice, Ting}, the tunneling junctions would serve as Cooper pair filters which divide Cooper pairs into two different groups. However all Cooper pairs occupy in the same macroscopic quantum state and there is no difference among them. It will break the basic principle of quantum mechanics to label the different internal relative moments ${\bf k}$ on Cooper pairs.  

Interference effects can measure the potential of the surrounding on the particles only \cite{Aharonov-Bohm, Berry}. As shown in Table~\ref{tab:table1}, interference effects depend on the entirety wavefunction of the particles and the internal structure is not related to it.  Particularly, in the two-photon EPR experiments \cite{two-photon}, the wave packets of the two paired photons are totally isolated, but the interference effect of photon pairs  can still be observed.

Therefore we conclude that the interference experiments can not  tell us anything about the pairing symmetry.

Now it is natural to ask the question:  Where does the $\pi$-phase shift come from? In the following, we propose a possible explanation. 

Superconductor is a macroscopic quantum system. The behavior of all Cooper pairs is the same as that of one pair because all condensed bosons occupy the same ground state.  Similar to the ordinary electron system, Cooper pairs also endure the effect of the periodic lattice potential $U({\bf R})$ including both electric and magnetic. Its entirety wavefunction obeys Bloch theory and is periodically modulated by $U({\bf R})$
\begin{equation}
 \Psi({\bf R})=u_{\bf K}({\bf R}) exp(i{\bf K} {\bf R})
\label{Eq:Psik0}
\end{equation}
where $u_{\bf K}({\bf R})$ is a periodic function with the same period as $U({\bf R})$. Without current, all Cooper pairs occupy the ${\bf K}=0$ state, so that the wavefunction can be expressed as
\begin{equation}
\Psi({\bf R})=u_0({\bf R}) 
\end{equation}

\begin{figure}
\includegraphics{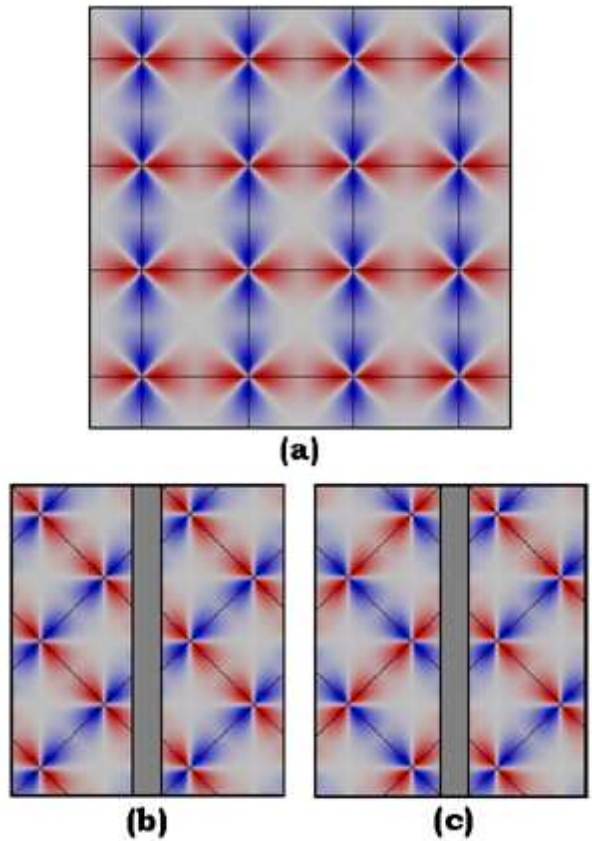}
\caption{\label{fig:Bloch} (a)Sketch map of Bloch-type superconducting condensate wavefunction $\Psi({\bf R})$. Red and blue denote different signs of the wavefunction and the crystalline lattice is denoted by the black lines. $\Psi({\bf R})$ changes sign within a crystalline cell. The white lines in 45$^{\circ}$ direction through lattice sites are nodal lines with $\Psi({\bf R})=0$. (b)(c)45$^{\circ}$ superconducting Josephson junctions with different wavefunction overlap}
\end{figure}

All cuprate superconductors have similar $CuO_2$ planes which are considered to be responsible for superconductivity. To make the discussion simple and clear, we consider the $CuO_2$ plane with strict 4/mmm point group symmetry as in $Tl_2Ba_2CaCu_2O_{6+\delta}$. In this case, the condensate wavefunction $\Psi({\bf R})$ keeps its amplitude under the mirror operators $M_a$, $M_b$ and $M_c$, and 90$^{\circ}$ rotation operator $C_4$, i.e. the possible eigenvalues for these symmetric operators are limited to be $\pm 1$. And the phase-sensitive experiments have proved that $C_4$ should be $-1$. Nonzero Josephson tunneling along a-axis and b-axis with a conventional superconductor requires that the eigenvalues for $M_a$, $M_b$ and $M_c$ should be $+1$. Then we have $M_i \Psi({\bf R})=  \Psi({\bf R})     (i= a, b, c)$, and $C_4 \Psi({\bf R})= -\Psi({\bf R})$. To meet these conditions, $\Psi({\bf R})$ should be zero along all 45$^{\circ}$ lines through lattice sites, as shown in Fig.~\ref{fig:Bloch}(a). The Bloch-type superconducting wavefunction around the lattice sites is very similar to the 3d$_{x^2-y^2}$ single electron wavefunction. This is of no surprise.  The superconducting condensate is made up with conduction electrons. The condensate wavefunction  $\Psi({\bf R})$ meets the following relationship 
\begin{equation}
|\Psi({\bf R})|^2 \propto n_p({\bf R}) \leq n_t({\bf R}),
\end{equation}
where $n_p({\bf R})$ and $n_t({\bf R})$ are  the density of paired electrons  and total conduction electrons respectively. 

The position-dependent Bloch-type superconducting wavefunction originates from the strong interaction between Cooper pairs and the lattice background. It should also exist in conventional metal superconductors. However, in conventional superconductors the periodic potential is rather smooth due to the relatively smaller lattice constant and higher electron density,  and therefore,  the modulation is not strong enough to change the sign of the superconducting condensate wavefunction.

In fact, there are already some experiments showing that the superconducting condensate wavefunction can be modulated by local background potentials, such as the amplitude change and even sign oscillation of the condensate wavefunction within the coherence length at the  superconductor/ferromagnet/superconductor junctions \cite{Chien}.

The Josephson coupling  between two superconductors resembles to the chemical bond between two atoms. There is a free energy change caused by  the overlap of the two superconducting condensate wavefunctions in the junction area
\begin{equation}
\Delta F= -c\int \Psi_L^* ({\bf R}) \Psi_R ({\bf R}) d{\bf R},
\label{Eq:F}
\end{equation}
where $c$ is a positive constant.  $\Delta F$ is always negative, so that the contribution from the region with the same sign is always  larger than that from the region with opposite signs. Therefore all the phase sensitive experiments \cite{TsueiRV} can be understood in the picture. 

For the junction configuration shown in Fig.~\ref{fig:paradox}c, the Josephson coupling strongly depends on the relative position of the two lattices. Fig.~\ref{fig:Bloch}(b) and (c) give the examples of two extreme relative positions.  The condensate wavefunction is anti-symmetric about the junction plane in Fig.~\ref{fig:Bloch}(b) (omit some microscopic mismatch) whereas it is symmetric in Fig.~\ref{fig:Bloch}(c). With continuously shifting the relative position from the former to the latter case, the coupling strength will first decrease, go to zero at a certain position, and then increase again. The condensate wavefunction is anti-symmetric about the junction plane before the position with zero coupling, and becomes symmetric after that position. With the relative position in Fig.~\ref{fig:Bloch}(b), when the thickness of the barrier approaches  zero, the system is reduced into a single superconductor and the wavefunction becomes that for one superconductor. Then, the paradox appeared in direction dependent picture\cite{Geshkenbein, Rice} no longer exists. In principle, the position dependence of Josephson tunneling along the $45^{\circ}$ direction can be tested by phase-sensitive experiments, though it is technically difficult. 

The above proposed picture can also be tested with the following observable experiment.  If one makes a junction by putting a tip of conventional superconductor to an a-b surface of a cuprate, along the 45$^{\circ}$ lines through the lattice sites, the critical current of the junction will be exactly zero because of the symmetry of the superconducting wavefunction. When the tip is placed at other positions, the critical current of the junction would be nonzero. Equidistantly scanning the tip above the a-b surface of a cuprate, one will obtain a map of critical current, and also an approximate map of amplitude of the Bloch-type superconducting wavefunction.

In conclusion, to describe a superconducting state, two wavefunctions are needed, condensate wavefunction and pairing wavefunction. The former describes the entirety movement of the pairs and the latter describes the relative movement of the two electrons within a pair. The superconducting interference effects show the properties of the condensate wavefunction, and do not depend on the pairing wavefunction. The $\pi$-phase shift observed in the phase sensitive Josephson measurements can not prove that the pairing state is d-wave. In stead, it tells us that the condensate wavefunction of Cooper pairs is a position dependent Bloch type wavefunction which changes its sign within a crystalline cell. Such a wavefunction is resulted from the interaction between cooper pairs and the positive periodic potential. Till now, it is still an open question whether the pairing state is d-wave or s-wave. Since the superconducting wavefunction strongly depends on the local potential, we expect to artificially modulate the phase and the sign of the wavefunction by changing the local potential via the methods such as local doping and hetero epitaxy, etc.  It is also possible, by using the local phase shift, to develop some quantum computing elements \cite{quantum-computer}.


\begin{references} 

\bibitem{Levi} B. G. Levi,  Phys. Today {\bf 46}, 17 (May 1993); {\bf 49}, 19 (Jan. 1996); {\bf 50}, 19 (Nov. 1997)

\bibitem{Anderson} P. W. Anderson,   Phys. Today {\bf 47}, 11 (Feb. 1994)

\bibitem{Chakravarty} S. Chakravarty,   Science {\bf 266}, 386 (1994)

\bibitem{Cox} D. L. Cox and M. B. Maple,  Phys. Today {\bf 48}, 32 (Feb. 1995)

\bibitem{Clery} D. Clery,  Science {\bf 271}, 288 (1996)

\bibitem{Kirtley} J. R. Kirtley and C. C. Tsuei,  Sci. Am. {\bf 95}, 50 (Aug. 1996)

\bibitem{TsueiRV}  C. C. Tsuei and J. R. Kirtley,  Rev. Mod. Phys. {\bf 72}, 969 (2000) and references therein

\bibitem{VanHarlingenRV} D. J. Van Harlingen,  Rev. Mod. Phys. {\bf 67}, 515 (1995) 

\bibitem{SigristRV} M. Sigrist and T. M. Rice,  Rev. Mod. Phys. {\bf 67}, 503 (1995)

\bibitem{ScalapinoRV} D. J. Scalapino,  Phys. Rep. {\bf 250}, 330 (1995)

\bibitem{Geshkenbein} V. B. Geshkenbein, A. I. Larkin, and A. Barone,  Phy. Rev. B. {\bf 36}, 235 (1987)

\bibitem{Rice} M. Sigrist and T. M. Rice,  J. Phys. Soc. Jpn. {\bf 61}, 4283 (1992)

\bibitem{two-photon} P. G. Kwiat, A. M. Steinberg, and R. Y. Chiao,  Phys. Rev. A {\bf 47}, R2472 (1993)

\bibitem{EPR} A. Einstein, B. Podolsky, and N. Rosen,  Phys. Rev. {\bf 47}, 777 (1935)  

\bibitem{NaBEC} M. R. Andrews, C. G. Townsend, H. -J. Miesner, D. S. Durfee, D. M. Kurn, W. Ketterle,  Science {\bf 275}, 637 (1997)

\bibitem{Ting} J. H. Xu, J. L. Shen, J. H. Miller, and C. S. Ting,  Phys. Rev. Lett. {\bf 73}, 2492 (1994)

\bibitem{Aharonov-Bohm} Y. Aharonov and D. Bohm,  Phys. Rev. 115, 485 (1959)

\bibitem{Berry} M. V. Berry,  Proc. R. Soc. London A {\bf 392}, 45 (1984). 

\bibitem{Chien} J.S.Jiang, D. Davidovi\'c, D. H. Reich, and C. L. Chien, Phys. Rev. Lett. {\bf 74}, 314 (1995)

\bibitem{quantum-computer} L. B. Ioffe, V. B. Geshkenbein, M. V. Feigel'man, A. L. Fauch\'ere, G. Blatter, Nature {\bf 398}, 679 (1999)

\bibitem{acknowledge} We acknowledge helpful discussion with Prof. D. L. Feng, Prof. S. P. Zhao, Prof. H. H. Wen, and Prof. N. L. Wang. 

\end{references}
\end{document}